\documentclass[aps,prd,preprint,nofootinbib,floatfix]{revtex4-1}
\usepackage[latin1]{inputenc}
\usepackage[T1]{fontenc}
\usepackage{graphicx}
\usepackage{amsmath}
\usepackage{amsfonts}
\usepackage{amssymb}
\usepackage{hyperref}
\usepackage{slashed}
\usepackage{feynmp}
\usepackage{feynmp-auto}
\usepackage{graphicx}
\usepackage[labelfont=bf,labelsep=quad,justification=justified]{caption}
\usepackage[latin1]{inputenc}
\usepackage{latexsym,amssymb} 
\usepackage{hyperref}
\usepackage{epstopdf}
\usepackage{graphicx} 
\usepackage[english]{babel}
\usepackage{tabularx}
\usepackage{listings} 
\usepackage{enumerate} 
\usepackage{fancyhdr} 
\usepackage{pstricks} 
\usepackage{color}
\usepackage{subfig} 
\usepackage{verbatim} 
\newcommand{\lsim}{\mathrel{\mathop{\kern 0pt \rlap
  {\raise.2ex\hbox{$<$}}}
  \lower.9ex\hbox{\kern-.190em $\sim$}}}
\newcommand{\gsim}{\mathrel{\mathop{\kern 0pt \rlap
  {\raise.2ex\hbox{$>$}}}
  \lower.9ex\hbox{\kern-.190em $\sim$}}}

\def  \bcen   {\begin{center}}
\def  \ecen   {\end{center}}
\def  \beq    {\begin{equation}}
\def  \eeq    {\end{equation}}
\def  \beqa   {\begin{eqnarray}}
\def  \eeqa   {\end{eqnarray}}

\def\bea{\begin{eqnarray}}
\def\eea{\end{eqnarray}}

\begin{document}
\title{Lepton number violation in  heavy Higgs decays to sneutrinos}

\author{S. Moretti$^{1,2}$}\email{S.Moretti@soton.ac.uk}
\author{C.H. Shepherd-Themistocleous$^{1,2}$}\email{Claire.Shepherd@stfc.ac.uk}
\author{H. Waltari$^{1,2,3}$}\email{H.Waltari@soton.ac.uk}

\affiliation{$^1$Particle Physics Department, STFC Rutherford Appleton Laboratory, Chilton, Didcot, Oxon OX11 0QX, UK}
\affiliation{$^2$School of Physics and Astronomy, University of Southampton, Highfield, Southampton SO17 1BJ, UK}
\affiliation{$^3$Department of Physics and Helsinki Institute of Physics, University of Helsinki, 00014 Helsinki, Finland}

\date{\today}

\begin{abstract}
\noindent We study the possibility of observing lepton number violation in the right-handed sneutrino sector of the Next-to-Minimal Supersymmetric Standard Model extended with right-handed neutrinos. The scalar potential introduces a lepton number violating mass term for the right-handed sneutrinos, which generates a phase difference that results in oscillations between the sneutrino and antisneutrino. If we have light higgsinos and right-handed sneutrinos, the sneutrino decay width is determined by the tiny Yukawa couplings, which allows the phase difference to accumulate before the sneutrino decays. We investigate the possibilities of producing sneutrino pairs resonantly through a heavy Higgs of such a model and the ability of seeing a lepton number violating signature emerging from sneutrinos at the Large Hadron Collider. We also discuss how a possible future signal of this type could be used to determine the neutrino Yukawa couplings.
\end{abstract}
\maketitle
\section{Introduction}

Neutrino oscillations \cite{Athanassopoulos:1997pv,Fukuda:1998mi,Aguilar:2001ty,Ahn:2002up,Abe:2011sj,An:2012eh} have shown us that flavor violation exists also in the leptonic sector of the Standard Model (SM). Neutrino oscillations imply non-zero neutrino masses, but these masses are so tiny that it seems unlikely that they are generated in the same manner as the masses of other SM fermions.

Extending the SM with $d=5$ terms leads to the so called Weinberg operator \cite{Weinberg:1979sa}, which does generate neutrino masses after Electro-Weak Symmetry Breaking (EWSB). This operator may be a remnant of a seesaw mechanism \cite{Minkowski:1977sc,Konetschny:1977bn,Mohapatra:1979ia,Magg:1980ut,Schechter:1980gr,Foot:1988aq} after one or several heavy particles have been integrated out. If the neutrino masses are generated via a seesaw mechanism, neutrinos are Majorana fermions and their mass term violates lepton number by two units ($\Delta L=2$). We may then ask whether we could see also lepton number violation so that we could get a confirmation of the Majorana nature of neutrinos. The smoking gun signature for this would be the neutrinoless double $\beta$ decay \cite{Schechter:1981bd} but so far experiments have not shown a clear signal of such a dynamics.

The SM has also other problems than neutrino mass generation so we need to extend it. Probably the most popular framework for doing so is Supersymmetry (SUSY), where the Poincar\'e group is extended by anticommuting generators, which lead to a symmetry between fermions and bosons. In order to avoid a fast proton decay one needs to impose $R$-parity, which leads to the lightest $R$-odd particle being stable and, if it is neutral, it would be a good Dark Matter (DM) candidate. In addition, an EW scale Higgs boson is natural in Supersymmetric models as the quartic coefficient of the Higgs potential arises from gauge couplings and the quadratic corrections to the Higgs boson mass stemming from boson and fermion loops cancel each other.

SUSY opens up the possibility of having lepton number violation through Superpartners. It is known that also sneutrinos have a Majorana character giving rise to lepton number violation \cite{Hirsch:1997is} but for left-handed sneutrinos this contribution needs to be tiny so that it would not generate a too large loop-induced contribution to neutrino masses \cite{Grossman:1997is}. Hence observing sneutrino-antisneutrino oscillations requires an extremely small decay width for the sneutrino, which is only possible in some very compressed scenarios \cite{Choi:2001fka}. The one-loop contribution involving right-handed sneutrinos is a lot smaller than for left-handed sneutrinos and hence the constraints from neutrino masses are not as severe \cite{Elsayed:2012ec}. In addition, the decay width of right-handed sneutrinos may be small due to the absence of gauge couplings so that the condition for lepton number violation in sneutrino decays $\Delta m_{\tilde{\nu}}/\Gamma_{\tilde{\nu}}\gtrsim 1$ \cite{Grossman:1997is} can be satisfied.

We show that in the Next-to-Minimal Supersymmetric Standard Model (NMSSM) extended with a type-I seesaw we may see a lepton number violating signal in the decays of the heavy CP-even or odd doublet Higgs state to sneutrinos. As we show, the heavy Higgs has even in the alignment limit\footnote{In the alignment limit one of the mass eigenstates is aligned in field space with the Vacuum Expectation Value (VEV) of the doublet states. For a discussion of the alignment limit in the NMSSM, see \cite{Carena:2015moc}.} a potentially large coupling to right-handed sneutrinos. Since in this model right-handed neutrinos and higgsinos get their masses through the same mechanism, we can expect the higgsinos to be roughly degenerate with the right-handed neutrinos. Soft SUSY breaking mass terms then make right-handed sneutrinos heavier than higgsinos. In such a case the right-handed sneutrino can decay visibly through its Yukawa interactions to a charged lepton and a chargino. If the heavy Higgs is not too heavy so that its production cross section is not too small and as long as the decay to right-handed sneutrinos is kinematically allowed, we could see the heavy Higgs decay via a same-sign dilepton signature together with soft jets and missing transverse energy. 
Furthermore, 
since the right-handed sneutrino is inert with respect to the gaugino component of the chargino, the decays would give us direct experimental access to the neutrino Yukawa couplings.

The plan of the paper is as follows. In the next section, we describe the NMSSM with right-handed neutrinos. Sect. III introduces lepton number violation in such a scenario. Sect. IV illustrates the phenomenology of heavy CP-even Higgs production and decay herein. Sect. V discusses our hallmark signature. In Sect. VI we explain how to access the neutrino Yukawa couplings while we conclude in the last section. 

\section{The NMSSM with right-handed neutrinos}

The NMSSM adds a gauge singlet chiral Superfield $S$ to the MSSM particle content and imposes a $\mathbb{Z}_{3}$ symmetry, which forbids the $\mu$-term. Once the scalar component of the singlet gets a  VEV, an effective $\mu$-term is generated, which has the right value to enable EWSB.

The NMSSM still lacks a mechanism for neutrino mass generation. Adding right-handed neutrinos allows us to introduce the type-I seesaw mechanism to explain small neutrino masses. This leads to the Superpotential \cite{Kitano:1999qb,Cerdeno:2008ep}
\begin{multline}\label{eq:superpotential}
W=y^{u}_{ij}(Q_{i}\cdot H_{u})U^{c}_{j}-y^{d}_{ij}(Q_{i}\cdot H_{d})D^{c}_{j}-y^{\ell}_{ij}(L_{i}\cdot H_{d})E^{c}_{j}+y^{\nu}_{ij}(L_{i}\cdot H_{u})N^{c}_{i}\\
+\lambda S(H_{u}\cdot H_{d})+\frac{\lambda_{Ni}}{2}SN_{i}^{c}N_{i}^{c}+\frac{\kappa}{3} S^{3},
\end{multline}
where repeated indices are summed over and we have introduced $A\cdot B\equiv \epsilon_{ab}A_{a}B_{b}$. As the VEV of the scalar component of the singlet Superfield $S$ generates also the right-handed neutrino mass term and we expect $|\mu_{\mathrm{eff}}|$ to be not too far above the EW scale, the neutrino Yukawa couplings need to be $\mathcal{O}(10^{-6})$ or smaller so that the neutrino mass $\sim (y^{\nu}v)^{2}/\lambda_{N}v_{s}$ would be in the sub-eV domain.

The VEVs of the scalar fields break the $\mathbb{Z}_{3}$ symmetry spontaneously leading to a potential problem with domain walls \cite{Zeldovich:1974uw}. The symmetry will be broken by higher dimensional operators, so that a preferred vacuum will exist, but this comes at the expense of destabilizing the hieararchy by generating soft SUSY breaking tadpole term $\xi^{3}S$ for the singlet, where $\xi$ is naturally at the scale of these non-renormalizable terms \cite{Abel:1995wk}. It is possible to impose an additional discrete symmetry that will be broken only by the soft SUSY breaking terms \cite{Panagiotakopoulos:1998yw} so that the coefficient $\xi$ of the singlet tadpole term will be of the order $m_{\mathrm{SUSY}}$, possibly suppressed by  loop factors. Also inflation may solve the domain wall problem \cite{Mazumdar:2015dwd}.

We write the soft SUSY breaking Lagrangian as
\begin{multline}
-\mathcal{L}_{\mathrm{soft}}=m^{2}_{\tilde{Q}}\tilde{Q}^{\dagger}\tilde{Q}+m^{2}_{\tilde{U}}|\tilde{U}|^{2}+m^{2}_{\tilde{D}}|\tilde{D}|^{2}+m^{2}_{\tilde{L}}\tilde{L}^{\dagger}\tilde{L}+m^{2}_{\tilde{E}}|\tilde{E}|^{2}+m^{2}_{\tilde{N}}|\tilde{N}|^{2}\\
+m_{S}^{2}|S|^{2}+m_{H_{u}}^{2}H_{u}^{\dagger}H_{u}+m_{H_{d}}^{2}H_{d}^{\dagger}H_{d}+M_{1}\tilde{B}\tilde{B}+M_{2}\tilde {W}\tilde{W}+M_{3}\tilde{g}\tilde{g}\\
+A^{u}_{ij}(\tilde{Q}_{i}\cdot H_{u})\tilde{u}^{*}_{j}-A^{d}_{ij}(\tilde{Q}_{i}\cdot H_{d})\tilde{d}^{*}_{j}-A^{\ell}_{ij}(\tilde{L}_{i}\cdot H_{d})\tilde{e}^{*}_{j}+A^{\nu}_{ij}(\tilde{L}_{i}\cdot H_{u})\tilde{N}^{*}_{j}\\
+A_{\lambda}S(H_{u}\cdot H_{d})+\frac{1}{2}A_{\lambda_{N}}S\tilde{N}_{i}\tilde{N}_{i}+\frac{1}{3}A_{\kappa}S^{3}+\xi^{3}S+\mathrm{h.c.},
\end{multline}
where we have included a term linear in $S$ arising from the non-renormalizable terms and $\xi$ being of the order of the SUSY breaking scale, as discussed.

\begin{figure}
\begin{center}
\includegraphics[width=0.9\textwidth]{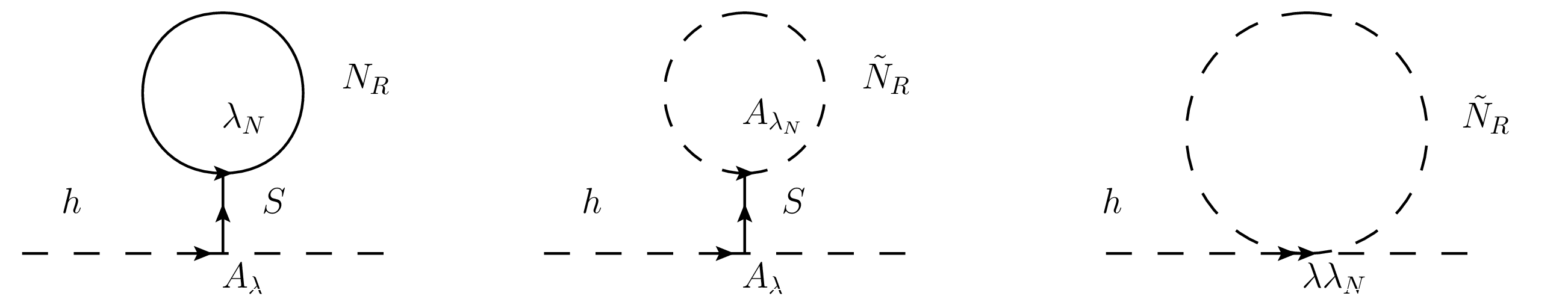}
\end{center}
\caption{Loops of right-handed neutrinos and sneutrinos can contribute to the SM-like Higgs mass if the couplings $\lambda$, $A_{\lambda}$ and $\lambda_{N}$ are large and the singlet Higgs is not too heavy. If these couplings are large, they can induce a large coupling between the heavy Higgs states and the right-handed sneutrinos. \label{fig:neutrinoloop}}
\end{figure}

The Superpotential (\ref{eq:superpotential}) leads to the following scalar potential (written here for one generation)
\begin{equation}
V=\left| \lambda H_{u}H_{d}+\frac{\lambda_{N}}{2}N^{2}+\kappa S^{2}\right|^{2}+\ldots,
\end{equation}
where we have explicitly written only the terms that are most interesting for lepton number violation in the sneutrino sector. The cross terms $\frac{\lambda\lambda_{N}}{2}H_{u}H_{d}N^{*}N^{*}+\frac{\lambda_{N}\kappa}{2}SSN^{*}N^{*}+\mathrm{h.c.}$ violate lepton number by two units and after EWSB lead to a mass difference between CP-even and CP-odd right-handed sneutrinos \cite{Cerdeno:2008ep,Cerdeno:2009dv}. The coupling to doublet Higgses is of interest as it can also be responsible for the annihilation of right-handed sneutrino DM so that the constraints from relic density can be satisfied \cite{Cerdeno:2008ep,Cerdeno:2009dv,Cerdeno:2011qv,Cao:2018iyk}. In addition, this term can lead to an increase of the SM-like Higgs mass through the neutrino-sneutrino loops of Figure \ref{fig:neutrinoloop} \cite{Wang:2013jya}, which does motivate the scenario for large $\lambda$ and $\lambda_{N}$. As we shall see, this part of the parameter space can result in a lepton number violating signature from sneutrino production.

\section{Sneutrinos in the presence of lepton number violation}

Sneutrinos have two bases of eigenstates, those given by the lepton number (sneutrino and antisneutrino, $\tilde{\nu}$ and $\tilde{\nu}^{*}$) or those given by the CP of the state (CP-even $\propto \tilde{\nu}+\tilde{\nu}^{*}$ and CP-odd $\propto \tilde{\nu}-\tilde{\nu}^{*}$, the real and imaginary parts of the sneutrino field). If lepton number is conserved, the CP-even and CP-odd fields propagate coherently together and are mass degenerate. If there is a lepton number violating term in the Lagrangian, it will generate a mass difference between the CP eigenstates and the accumulated phase difference can lead to observable lepton number violation. In order to have an observable signature, the phase difference needs to be large enough before the sneutrino decays. The amount of visible lepton number violation can be parametrized by $x=\Delta m_{\tilde{\nu}}/\Gamma_{\tilde{\nu}}$. The probability for a sneutrino to oscillate to an antisneutrino is given by \cite{Ghosh:2010ns}
\begin{equation}
P(L)=\frac{e^{-\alpha L}}{2(\alpha^{2}+\beta^{2})}\left[ -\alpha^{2}+(e^{\alpha L}-1)\beta^{2}+\alpha^{2}\cos(\beta L)-\alpha\beta \sin(\beta L) \right],
\end{equation}
where $\alpha=\Gamma m/E$, $\beta=\Delta m^{2}/2E$ and $L$ is the distance the sneutrino has travelled. In the limit of large $L$ and $x$ this reduces to $P(L)=\frac{x^{2}}{2(1+x^{2})}\rightarrow 1/2$ as we may expect.

If the slepton sector is CP-conserving, the sneutrino mass matrix for one generation can be given in the basis $(\Re(\tilde{\nu}_{L}),\Re (\tilde{N}_{R}),\Im(\tilde{\nu}_{L}), \Im (\tilde{N}_{R}) )$ as \cite{Cerdeno:2009dv}
\begin{equation}
\mathcal{M}^{2}_{\tilde{\nu}}=
\begin{pmatrix}
m_{LL}^{2} & m_{LR}^{2}+m_{RL}^{2} & 0 & 0\\
m_{LR}^{2}+m_{RL}^{2} & M_{RR}^{2}+m_{RR}^{2} & 0 & 0\\
0 & 0 & m_{LL}^{2} & -m_{LR}^{2}+m_{RL}^{2}\\
0 & 0 & -m_{LR}^{2}+m_{RL}^{2} & M_{RR}^{2}-m_{RR}^{2}
\end{pmatrix},
\end{equation}
where
\begin{eqnarray}
m_{LL}^{2} & = & m_{\tilde{L}}^{2}+|y^{\nu}v_{u}|^{2}+\frac{1}{2}m_{Z}^{2}\cos 2\beta,\\
m_{LR}^{2} & = & y^{\nu}A_{\nu}v_{u}-y^{\nu}\lambda v_{s}v_{d},\\
m_{RL}^{2} & = & -y^{\nu}\lambda v_{s}v_{u},\\
M_{RR}^{2} & = & m_{\tilde{N}}^{2}+|y^{\nu}v_{u}|^{2}+|\lambda_{N}v_{s}|^{2}\quad \mathrm{and}\\
m_{RR}^{2} & = & A_{\lambda_{N}}v_{s}+\lambda_{N}\kappa v_{s}^{2}-\lambda\lambda_{N}v_{u}v_{d}.\label{eq:lnvmass}
\end{eqnarray}
Here $m_{RR}^{2}$ is the lepton number violating mass term that generates the mass difference between the CP eigenstates. We may note that  in this particular model the left-handed sneutrinos have no lepton number violating mass terms.

The lepton number violating term induces a loop contribution to the left-handed neutrino masses \cite{Grossman:1997is}. At one loop the contribution from right-handed sneutrinos is proportional to $(y^{\nu})^{2}$ and hence negligible \cite{Elsayed:2012ec} and also the two-loop contribution is suppressed in this model. At the three-loop level there are diagrams which are not suppressed by powers of $y^{\nu}$. If we assume that we get a suppression about three orders of magnitude per loop\footnote{This is true for the leading one-loop contribution from left-handed sneutrinos \cite{Grossman:1997is} and we do expect that, by taking into account diagrams where bosons and fermions are interchanged, we should get further cancellations.}, we are safe with $\Delta m_{\tilde{\nu}}\lesssim 1$~GeV or slightly more. We may notice that for the NMSSM extended with other types of seesaw mechanisms the first unsuppressed contribution comes at one or two loops so the constraint on sneutrino mass differences is at most at the MeV level. In our case we do need some cancellations between the different terms of $m_{RR}^{2}$ to achieve $\Delta m_{\tilde{\nu}}\lesssim 1$~GeV, but we consider this to be less fine-tuned than other scenarios with sneutrino-antisneutrino oscillation.

\begin{figure}
\begin{center}
\includegraphics[width=\textwidth]{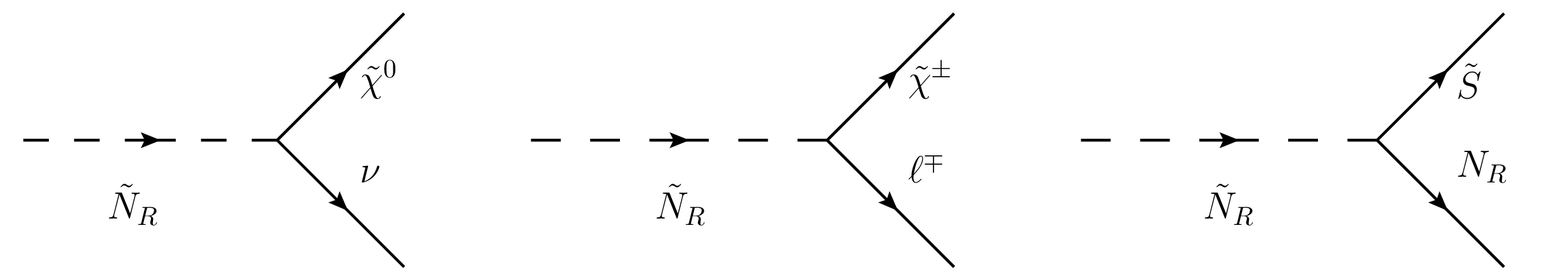}
\end{center}
\caption{Decay modes for the right-handed sneutrino. We are mostly interested in the second one as it has charged leptons in the final state.\label{fig:sneutrinodecay}}
\end{figure}

We explore the case where the gauginos are heavier than the right-handed sneutrinos, but the higgsinos are lighter. The higgsino masses arise from the same singlet VEV as the masses of right-handed neutrinos and hence higgsinos and right-handed neutrinos are naturally (\textit{i.e.} we assume $\lambda$ and $\lambda_{N}$ to be of similar size) at the same scale. We then expect the soft SUSY breaking masses to make the sneutrinos heavier than the higgsinos. The right-handed sneutrino has three possible decay modes
\begin{equation}
\tilde{N}_{R}\rightarrow \nu_{L}\tilde{H}^{0},\quad \tilde{N}_{R}\rightarrow \ell^{\pm}\tilde{H}^{\mp},\quad \tilde{N}_{R}\rightarrow \tilde{S}N_{R},
\end{equation}
depicted in Figure \ref{fig:sneutrinodecay}, of which the last one may not (and usually will not) be kinematically allowed. If the singlino is so heavy that the last decay is forbidden, the sneutrino decays are dictated by the neutrino Yukawa couplings and up to phase space effects, the Branching Ratios (BRs) of the two higgsino modes are equal. Since the Yukawa couplings are superpotential couplings, the non-renormalization theorems protect them from any large loop corrections. As the Yukawa couplings are very small, the decay width of the right-handed sneutrino is small, too, and the condition $\Delta m_{\tilde{N}}/\Gamma_{\tilde{N}} > 1$ can easily be satisfied. The decay widths are actually in the sub-eV regime, so one automatically has the case where probabilities for either sign leptons are $1/2$.

The decay $ \tilde{N}_{R}\rightarrow \ell^{\pm}\tilde{H}^{\mp}$ is especially interesting as it is visible. Once we have a lepton number violating mass term as in eq. (\ref{eq:lnvmass}) , the sneutrino that is created as a state with definite lepton number is not an eigenstate of the Hamiltonian but will start to oscillate between the sneutrino and antisneutrino states. In the limit $\Delta m_{\tilde{N}}\gg \Gamma_{\tilde{N}}$ it will quickly relax to one of the CP eigenstates, which do not have a well-defined lepton number so the decays may be lepton number violating. If  we produce a pair of sneutrinos and both of them decay visibly, we may get a same-sign dilepton signature, for which the SM background is a lot smaller than for opposite-sign dileptons. The chargino can decay either via $\tilde{\chi}^{\pm}\rightarrow \tilde{\chi}^{0}\ell^{\pm}\nu$ or $\tilde{\chi}^{\pm}\rightarrow \tilde{\chi}^{0}jj$ (hereafter, $\ell =e,\mu$ and $j$ represents a jet) of which the latter has the larger BR and it also allows us to see the same-sign dilepton final state instead of three leptons. If there is a third lepton in the final state from the decay of the chargino, it will be a lot softer than the leptons originating from the first step of the sneutrino decay as the mass gap between the charged and neutral higgsino is small, in this model typically $5$--$10$~GeV. We shall veto against a third hard lepton to reduce the background from dibosons. We do expect some hadronic activity as in the majority of the events at least one of the charginos leaves (soft) jets in the final state.

We also note that the light right-handed sneutrinos give only a small contribution to lepton flavor violating observables, since they only couple to higgsinos with tiny Yukawa couplings. This would not be the case for left-handed sneutrinos, where their gauge interactions would require almost perfect flavor alignment to satisfy the bounds from $\mu\rightarrow e\gamma$ if both sneutrinos and gauginos are light.

\section{Heavy Higgs production and decay}

From the Higgs data collected by the Large Hadron Collider (LHC) we know that any Beyond the SM (BSM) Higgs scenario must be near the alignment limit, including our modified NMSSM. In the alignment limit the heavy Higgs couplings to vector bosons vanish. Hence  heavy Higgs states cannot be produced (in large numbers) through vector boson fusion or Higgs-strahlung. In our scenario, the couplings to SM fermions in the alignment limit are either enhanced (down-type quarks, charged leptons) or suppressed (up-type quarks) compared to a SM-like Higgs of the same mass, assuming $\tan \beta=v_{u}/v_{d} > 1$. This means that the gluon fusion production cross section is modified from the SM Higgs one, but can be reasonably large. Conversely, gluon fusion does not give any additional objects that would help in tagging heavy Higgs decays so that the decays to quarks are lost in the QCD background.

The gluon fusion cross section falls rather steeply with the Higgs mass and hence we are quite quickly limited by the heavy Higgs production rate. The rates for the signals we are interested in will not be distinguishable from the background before the High Luminosity LHC (HL-LHC) if the heavy Higgs is heavier than $500$--$600$~GeV.

If no SUSY modes are kinematically allowed the heavy Higgs decays mainly to third generation SM fermions, usually $H\rightarrow b\overline{b}$ dominates. Due to the QCD background the hadronic modes do not give the best limits, instead we have to rely on $H\rightarrow \tau^{+}\tau^{-}$ \cite{Aaboud:2017sjh,Sirunyan:2018zut}. This decay mode excludes non-SM Higgs bosons below $500$~GeV if $\tan\beta \gtrsim 10$ and masses beyond $1$~TeV for large values of $\tan\beta$. We pick rather moderate values for $\tan \beta$, which are still allowed by direct searches for heavy Higgs bosons for masses between $400$ and $500$~GeV.

The vector boson decay modes are absent in the alignment limit and the three-scalar coupling responsible for $H\rightarrow hh$ also vanishes in the exact alignment limit \cite{Grzadkowski:2018ohf}. It is however interesting that the couplings to right-handed sneutrinos (neglecting the part from Yukawa couplings) in the alignment limit are
\begin{eqnarray}
C_{h\tilde{N}\tilde{N}} & = & \pm\frac{1}{2}\lambda\lambda_{N}v\sin 2\beta,\\
C_{H\tilde{N}\tilde{N}} & = & \pm\frac{1}{2}\lambda\lambda_{N}v\cos 2\beta,
\end{eqnarray}
where the upper (lower) sign is for CP-even (CP-odd) sneutrinos. If $\tan \beta > 1.5$, the coupling to the heavy Higgs is larger and when $\tan\beta\gg 1$ we have $\cos2\beta\simeq -1$ and $\sin 2\beta$ small. Hence we expect the heavy Higgs to have a large coupling to the sneutrinos. This is due to the form of the scalar potential: the lepton number violating term mixes $H_{u}$ and $H_{d}$ so that $h$ gets replaced by its VEV in the four-point coupling while $H$ gets the larger three-point coupling. The BR$(H\rightarrow \tilde{N}\tilde{N})$ can be large, whenever both $\lambda$ and $\lambda_{N}$ are large. Numerically we can get BRs up to $8\%$ for $H\rightarrow \tilde{N}\tilde{N}$ for $\lambda, \lambda_{N}\lesssim 0.7$.

The alignment limit couplings will be modified by the mixing between the singlet and the heavy doublet states. The singlet gauge eigenstate and sneutrinos would have a coupling
\begin{equation}
C_{S\tilde{N}\tilde{N}} =  |\lambda_{N}|^{2}v_{S}\pm(\kappa\lambda_{N}v_{S}+A_{\lambda_{N}}).
\end{equation}
If $\lambda$ is taken positive we expect singlet-doublet mixing in the Higgs sector to result in a partial cancellation in the coupling to CP-odd sneutrinos and an enhancement to CP-even sneutrinos.

Also the heavy CP-odd Higgs has a similar coupling to a CP-even and a CP-odd sneutrino and, since it is roughly degenerate with the heavy CP-even Higgs, it also contributes to the production of sneutrino pairs. The BR for $A\rightarrow \tilde{N}_{s}\tilde{N}_{p}$ is usually somewhere between one third to a half of the BR$(H\rightarrow \tilde{N}_{s}\tilde{N}_{s})$, where we have noted the CP-even (scalar) and CP-odd (pseudoscalar) components of the sneutrino by $\tilde{N}_{s}$ and $\tilde{N}_{p}$, respectively.

We also note that the heavy Higgs has a chance of decaying to higgsino pairs, of which the charged ones will lead to leptonic signatures but only in the opposite-sign dilepton channel, where the SM background is larger. Hence we do not expect these modes to be easily discovered, but they ameliorate the constraints from heavy Higgs searches by making the BR to SM final states smaller.

\section{Lepton number violating Higgs decays at the LHC}

\subsection{Signal and SM backgrounds}

\begin{figure}
\begin{center}
\includegraphics[width=0.9\textwidth]{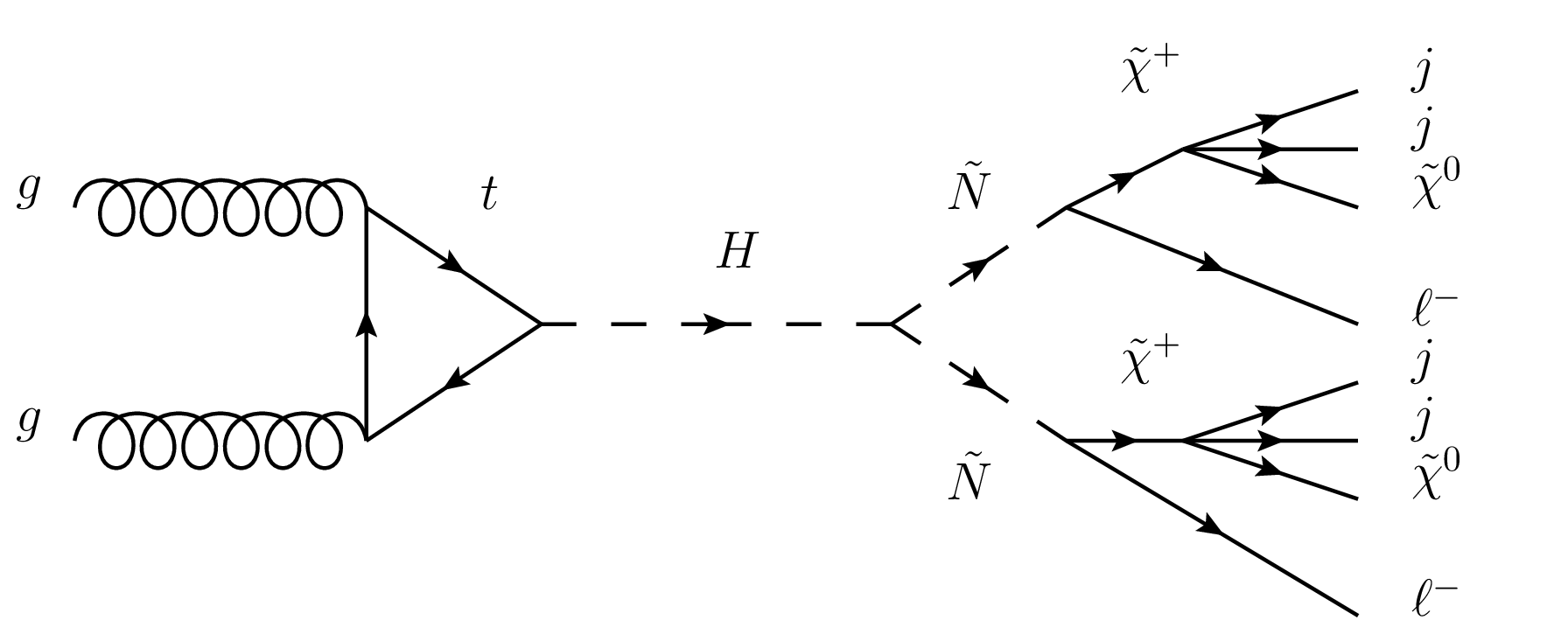}
\end{center}
\caption{An example of a full process leading to the same-sign dilepton + $\slashed{E}_{T}$ signature. There can also be a lepton and a neutrino ($\ell\nu$) instead of hadrons ($jj$) and similar diagrams for the CP-odd Higgs.\label{fig:decaychain}}
\end{figure}

We now consider the process $pp\rightarrow H, A \rightarrow \tilde{N}\tilde{N}$ and the subsequent decays $\tilde{N}\rightarrow \ell^{\pm}\tilde{\chi}^{\mp}$ at the LHC. We show an example of such a process in Figure \ref{fig:decaychain}. The overall production cross section depends mostly on $m_{H}$. Since $m_{H}> 2m_{\tilde{N}}>2m_{\tilde{\chi}}$ and the bounds on the higgsino masses are around $160$~GeV, the heavy Higgs needs to be beyond $400$~GeV to have any hope for a visible signal --- if there is no phase space available, the leptons will be so soft that they will  not pass trigger selection and also the missing transverse momentum will be small. To achieve this mass hierarchy and being compatible with other SUSY searches we have to assume that the soft SUSY breaking mass terms for right-handed sneutrinos are clearly smaller than other soft SUSY breaking masses.

We prepared the model files with the Mathematica package \textsc{Sarah v4.14} \cite{Staub:2008uz,Staub:2013tta}. Since $\Delta m_{\tilde{N}}/\Gamma_{\tilde{N}}\gg 1$ we simply used the CP-eigenstates of right-handed sneutrinos throughout while the eigenstates of definite lepton number were used for left-handed sneutrinos. The spectra were computed with \textsc{SPheno v4.0.3} \cite{Porod:2003um,Porod:2011nf}.

The signal we are searching for is two same-sign same-flavor leptons and missing transverse energy. A third hard lepton will be vetoed. The major SM backgrounds to this final state come from:
\begin{itemize} 
\item $WZ$ production in the case, where one lepton is not identified or has low $p_{T}$ (transverse momentum),
\item same-sign $WW$ production, where both $W$ bosons decay leptonically,
\item non-prompt leptons, which originate, \textit{e.g.}, from heavy-flavor mesons.
\end{itemize}

The non-prompt lepton background comes mainly from $t\overline{t}$ when one lepton comes from a $W$ boson and the other from the decay of a $B$ meson \cite{Chatrchyan:2011wba}. Also $W$+jets contributes to the non-prompt lepton background but the contribution is subdominant in the signal regions we are considering.

\subsection{Event selection and cuts}
We shall simulate the signal and background using the CMS detector as our example (see later on for details of its implementation). Our signal consists of two isolated\footnote{Isolation means a $\Delta R > 0.4$ separation to any other lepton and the lepton carrying at least $80\%$ ($90\%$) of the transverse momentum within a cone of $\Delta R <0.5$ in case of a muon (electron).} same-sign same-flavor leptons ($\ell=e,\mu$). The CMS dilepton trigger requires the leading electron (muon) to have $p_{T}>23 (17)$~GeV and the trailing electron (muon) $p_{T}>12 (8)$~GeV. We shall require $p_{T}>25$~GeV for the leading lepton and $p_{T}>12$~GeV for the second lepton as these reduce the background for non-prompt leptons, where often at least one of the leptons is rather soft. We shall veto a third lepton with $p_{T}>20$~GeV, which will reduce the $WZ$ background. We also veto against opposite-sign same flavor lepton pairs with an invariant mass compatible with the $Z$ boson (invariant mass between $80$ and $100$~GeV rejected).

\begin{table}
\begin{center}
\begin{tabular}{l c}
\hline
\hline
Dilepton trigger: & \\
$p_{T}(\ell_{1})_{\mathrm{min}}$ & $25$~GeV\\
$p_{T}(\ell_{2})_{\mathrm{min}}$ & $12$~GeV\\
$|\eta (\ell)|$ & < 2.5 \\
\hline
Same-sign same-flavor lepton pair: & \\
$N(e^{-})$ or $N(e^{+})$ or $N(\mu^{-})$ or $N(\mu^{+})$ & $2$\\
\hline
Veto for third hard lepton: $p_{T}(\ell_{3})$ & $<20$~GeV\\
Veto for $Z$ bosons: $M(\ell^{+}\ell^{-})$ & not $[80,100]$~GeV\\
\hline
\hline
\end{tabular}
\end{center}
\caption{The criteria used for event selection.\label{tb:selection}}
\end{table}

\begin{figure}
\begin{center}
\includegraphics[width=0.75\textwidth]{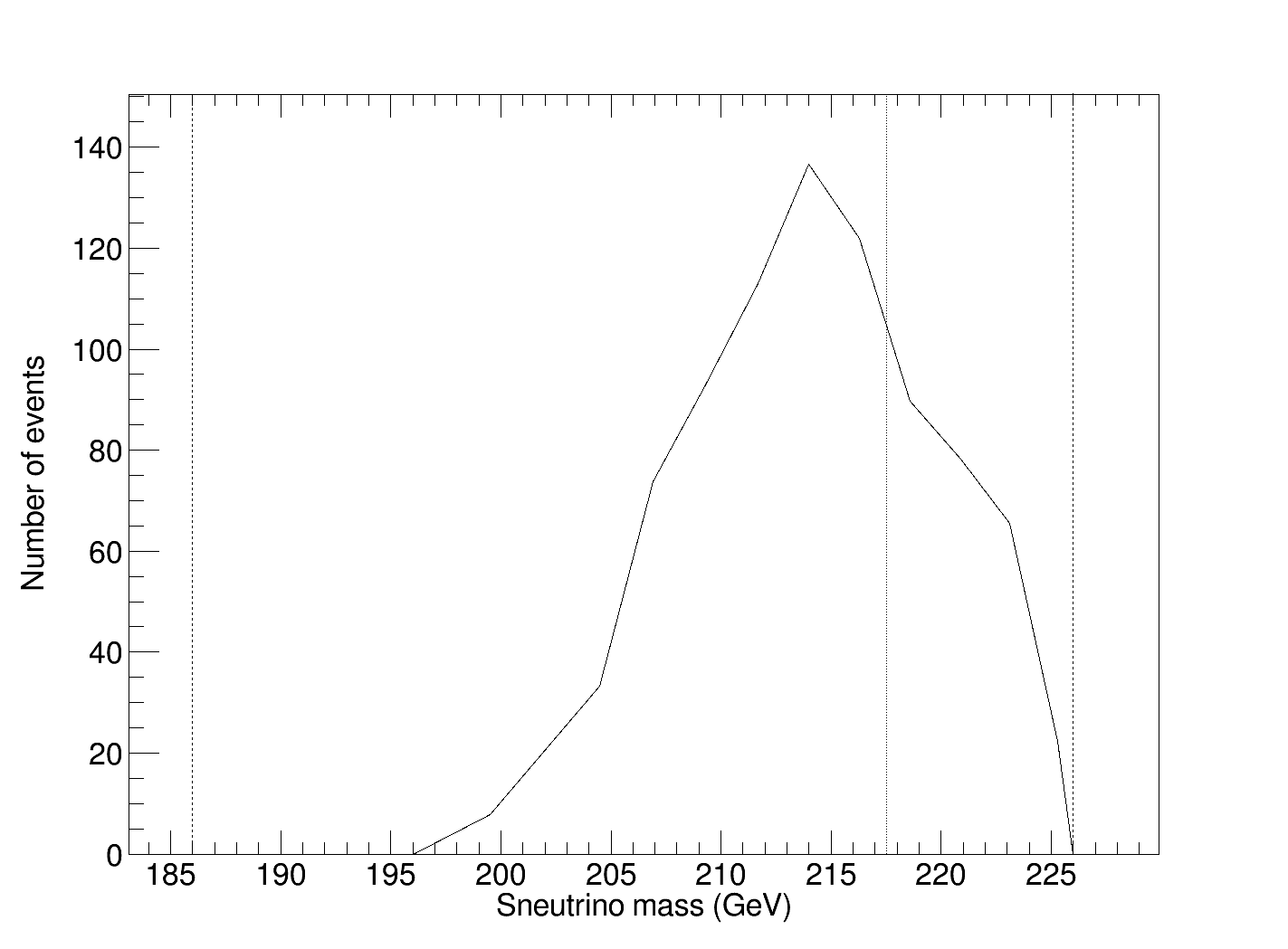}
\end{center}
\caption{We show the number of events passing the criteria of Table \ref{tb:selection} as a function of sneutrino mass for a benchmark with $\lambda=0.52$, $\lambda_{N}=0.50$, $\tan\beta=2.8$, $m_{\tilde{\chi}^{\pm}}=186$~GeV (first vertical line), $m_{H}=452$~GeV (last vertical line $m_{H}/2$) and $m_{A}=435$~GeV (intermediate vertical line $m_{A}/2$). Unless the sneutrino is at least $15$~GeV heavier than the chargino, the search based on isolated leptons is not efficient. The best sensitivity is at masses slightly below $m_{A}/2$.\label{fig:snumass}}
\end{figure}

Our event selection is not sensitive to the case where the sneutrino is only slightly heavier than the chargino. In such a case the leptons usually are not isolated and, if they are, their $p_{T}$ is so low that they will not survive the trigger cuts. We demonstrate this in Figure \ref{fig:snumass}. In what follows, we concentrate on the region of parameter space with $m_ {\tilde{N}}-m_{\tilde{\chi}^{\pm}}>15$~GeV, where the search based on isolated leptons can be efficient. We leave the case of a compressed spectrum for future work.

We propose two Signal Regions (SRs) by imposing cuts. These are partially overlapping. The first one targets the case where the spectrum is somewhat compressed whereas the second one is better when there is more phase space available.

Our first requirement is $\slashed{E}_{T}>50 (100)$~GeV for SR1 (SR2). We expect a reasonably large amount of missing transverse momentum due to the neutralinos in the final state of the signal while SM processes tend to have $\slashed{E}_{T}$ (missing transverse energy) close to $m_{W}/2$. The stronger cut on $\slashed{E}_{T}$  gives further suppression especially of the $WZ$ background.

We also require $10$~GeV $<M_{\ell_{1}\ell_{2}}< 50 (80)$~GeV for SR1 (SR2). The lower bound is imposed to reject leptons originating from $B$ mesons. The upper bound is chosen so that it captures most of the signal. As the leptons emerge from the heavy Higgs and the final state includes several other particles, the invariant mass of the lepton pair tends to be rather low. If the spectrum is compressed the invariant mass of the lepton pair is low and hence even the $50$~GeV cut accepts almost all of the signal. 

\begin{figure}
\begin{center}
\includegraphics[width=0.85\textwidth]{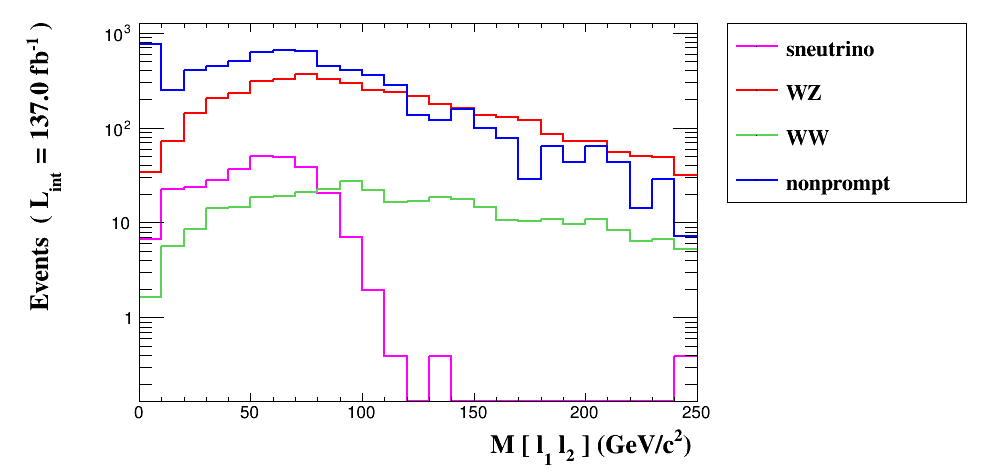}
\end{center}
\caption{The distribution of the invariant mass of the two leptons with highest $p_{T}$ for the signal (BP1, Table \ref{tb:benchmarks}) and  background components. We have applied the criteria of Table \ref{tb:selection}. We may see that the signal concentrates in the lower end of the invariant mass of the lepton pair. On the other hand there are a lot of background events from non-prompt leptons with very low invariant masses, hence we choose the signal regions to be between $10$ and $50$ or $80$~GeV. \label{fig:cut1}}
\end{figure}

The upper limit is chosen because, if a significant part of the signal would populate $M(\ell_{1} \ell_{2})>80$~GeV, the phase space needed would require the heavy Higgs to be so heavy that the production cross section is very low. The signal would be unobservable as the overall number of signal events would be low and lifting the upper limit on $M(\ell_{1} \ell_{2})$ would increase the background. We present the $M(\ell_{1}, \ell_{2})$ distributions of the background components and one signal benchmark in Figure \ref{fig:cut1}.

To suppress the background from $t\overline{t}$ we veto against $b$-jets. The $b$-jet veto means rejecting any object identified as a $b$-jet according to the loose identification scheme described in \cite{Chatrchyan:2012jua}. This does mean that we also loose a part of the signal (which contains basically no $b$-quarks due to CKM suppression) as especially $c$-quarks may get misidentified as $b$-quarks, but still about $75\%$ of the $c$-quarks and $90\%$ of light quarks or gluons pass the cut. In addition, due to the small mass gap between the chargino and neutralino, a large fraction of the hadrons will be with quite a low $p_{T}$ so some of them will not be reconstructed as jets.

\begin{figure}
\begin{center}
\includegraphics[width=0.85\textwidth]{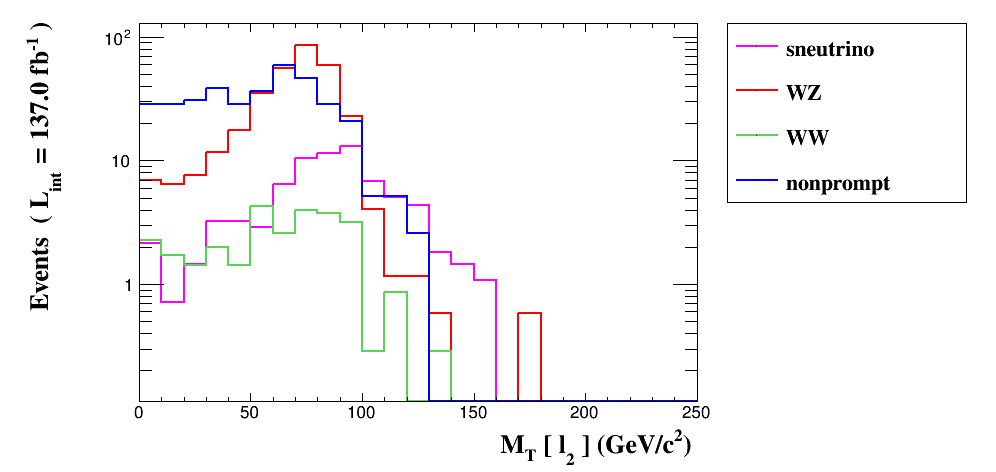}
\end{center}
\caption{The transverse mass of the lepton with second highest $p_{T}$ for the signal (BP1, Table \ref{tb:benchmarks}) and background components. We see that in the region $M_{T}(\ell_{2})>100$~GeV the signal is comparable to the background. In this plot all other cuts of SR1 given in Table \ref{tab:cuts} have been applied except the last cut on $M_{T}(\ell_{2})$. \label{fig:cut2}}
\end{figure}

We also impose a cut on $M_{T}(\ell_{2})=\sqrt{2\slashed{p}_{T}p_{T}(\ell_{2})(1-\cos(\Delta\phi))}$, where $\Delta\phi$ is the azimuthal angular separation between $\vec{p}(\ell_{2})$ and $\vec{\slashed{p}}$. This last cut is efficient against the background since the signal has two somewhat hard leptons of which one or the other might be more or less back-to-back against the total missing transverse momentum. Especially for non-prompt leptons the second lepton is typically softer and hence the transverse mass will be small. If the lepton is harder, the $W$ boson typically gets a recoil in the opposite direction in which case both the hardest lepton and the missing transverse momentum are more or less back-to-back to the second lepton. Such events are largely rejected by the cut on $M(\ell_{1} \ell_{2})$ and hence the higher end of the $M_{T}(\ell_{2})$ spectrum has rather small background contamination, while still containing a reasonable number of signal events as shown in Figure \ref{fig:cut2}. If no cut on the invariant mass of the lepton pair were imposed, the cut on $M_{T}(\ell_{2})$ would have a similar efficiency for both  signal and  background. A summary of the cuts is given in Table \ref{tab:cuts}.

\begin{table}
\begin{center}
\begin{tabular}{l c c}
\hline
\hline
Cut & SR1 & SR2\\
\hline
Missing transverse energy & & \\
$\slashed{E}_{T}$ & $>50$~GeV & $>100$~GeV\\
\hline
Lepton pair invariant mass & &\\
$M(\ell_{1}\ell_{2})$ & $>10$~GeV & $>10$~GeV\\
& $<50$ GeV & $<80$ GeV \\
\hline
Veto for b-jets: $N(b)$ & $0$ & $0$\\
\hline
Cut on second lepton $M_{T}$ & & \\
$M_{T}(\ell_{2})$ & $> 100$~GeV & $> 100$~GeV\\
\hline
\hline
\end{tabular}
\end{center}
\caption{The cuts that are used to define the signal regions. The cuts of Table \ref{tb:selection} are included to both signal regions as a preselection. These signal regions partially overlap. \label{tab:cuts}}
\end{table}

The nature of the signal and the composition of the background makes it hard to introduce cuts on hadronic activity beyond the $b$-jet veto. While $WW$ events are accompanied by two jets and non-prompt lepton events have typically hadronic activity, also the signal often has some hadronic activity from the chargino decays. Conversely, $WZ$ events have lower hadronic activity so their relative number is increased if we try to cut either on jet momenta or total hadronic activity.

\subsection{Benchmarks}

We have prepared a set of Benchmark Points (BPs). The experimental lower bound for the lightest higgsino is around $160$~GeV as the mass differences between the other higgsinos are in the range of $5$--$10$~GeV \cite{Aaboud:2017leg,Sirunyan:2018iwl}. We take the higgsino masses close to this lower bound and choose the soft SUSY breaking sneutrino masses so that the  lightest sneutrino is between $200$ and $250$~GeV and the heavy Higgses somewhat heavier than twice the sneutrino mass. For simplicity we choose the soft SUSY breaking sneutrino masses so that in most benchmarks there will be only one sneutrino that would be kinematically accessible.

We choose $\tan\beta$ to be between $2$ and $5$ so that the direct searches for heavy Higgses would not have excluded the BPs. We take $\lambda$ and $\lambda_{N}$ close to $0.5$, which lead to BRs around $5\%$ for $H\rightarrow \tilde{N}\tilde{N}$. We give the most important parameters for our BPs in Table \ref{tb:benchmarks}.

\begin{table}
\begin{tabular}{l c c c c c c}
\hline
\hline
Parameter & BP1 & BP2 & BP3 & BP4 & BP5 & BP6\\
\hline
Heavy CP-even Higgs mass& $455$ & $569$ & $484$ & $478$ & $490$ & $448$\\
Heavy CP-odd Higgs mass& $441$ & $562$ & $470$ & $464$ & $476$ & $434$ \\
Lightest sneutrino mass& $220$ & $233$ & $220$ & $227$ & $214$ & $210$ \\
Second sneutrino mass& $310$ & $321$ & $240$ & $338$ & $338$ & $340$\\
Chargino mass& $177$ & $190$ & $178$ & $178$ & $185$ & $186$ \\
$\lambda$ & $0.50$ & $0.50$ & $0.50$ & $0.50$ & $0.52$ & $0.52$ \\
$\lambda_{N}$ & $0.62$ & $0.62$ & $0.62$, $0.68$ & $0.64$ & $0.46$ & $0.46$ \\
$\tan\beta$ & $2.5$ & $4.0$ & $2.5$ & $2.5$ & $2.7$ & $2.7$\\
BR($H\rightarrow \tilde{N}\tilde{N}$)& $5.1\%$ & $5.0\%$ & $4.9\%+1.6\%$ & $4.7\%$ & $2.6\%$ & $3.7\%$ \\
BR($A\rightarrow \tilde{N}\tilde{N}$)& $0.5\%$ & $2.3\%$ & $1.3\%$ & $0.9\%$ & $1.1\%$ & $1.5\%$\\
BR($\tilde{N}\rightarrow e^{\pm}\tilde{\chi}^{\mp}$) & $40\%$  & $37\%$ & $40\%$, $1\%$ & $1\%$ & $1\%$ & $1\%$ \\
BR($\tilde{N}\rightarrow \mu^{\pm}\tilde{\chi}^{\mp}$) & $1\%$  & $1\%$ & $1\%$, $48\%$ & $47\%$ & $44\%$ & $42\%$ \\
BR($\tilde{N}\rightarrow \tau^{\pm}\tilde{\chi}^{\mp}$) & $8\%$  & $8\%$ & $8\%$, $1\%$ & $1\%$ & $1\%$ & $1\%$ \\
BR($\tilde{N}\rightarrow$ invisible) & $51\%$  & $54\%$ & $51\%$, $50\%$ & $51\%$ & $54\%$ & $56\%$ \\
\hline
\hline
\end{tabular}
\caption{The parameters for the signal BPs. The masses are given in units of GeV. In BP3 we have two sneutrinos the heavy Higgs can decay to, the first numbers refer to the lighter, the second ones to the heavier. For all other benchmarks the branching ratios refer to the lightest one, which is the only sneutrino that is kinematically accessible. \label{tb:benchmarks}}
\end{table}

For all benchmarks $m_{h}=125\pm 1$~GeV and all other Superpartners except right-handed sneutrinos and higgsinos are decoupled, \textit{i.e.}, at least heavier than the heavy Higgses.
BP1 has only a heavy Higgs at $455$~GeV and one sneutrino that is lighter than $m_{H}/2$. Since $m_{\tilde{N}}\simeq m_{A}/2$ basically only the CP-even Higgs contributes to the lepton number violating signal. The sneutrino decays mostly to electrons or positrons and charginos. BP2 is similar but with a heavier spectrum.
BP3 differs from these as the heavy Higgs can decay to two flavors of sneutrinos, the lighter decaying mostly to electrons and the heavier to muons. 
In BP4, BP5 and BP6 the lightest sneutrino decays dominantly to muons and again the Higgs has only one sneutrino flavor to decay to. These differ in the mass splittings between the heavy Higgs, the sneutrino and the chargino, BP4 having the largest splittings and BP6 being the most compressed spectrum with BP5 in between.

\subsection{Simulated results}

The diboson backgrounds are estimated by simulating them with \textsc{MadGraph5 v2.6.4} \cite{Alwall:2011uj} at Leading Order LO  and correcting the rates to Next-to-LO (NLO) accuracy with $K$-factors of $1.5$ for $WW$ production and $1.8$ for $WZ$ production (based on NLO computation with \textsc{MadGraph5} and the results of \cite{Grazzini:2016swo}). The background from non-prompt leptons is simulated by generating $t\overline{t}$ events, where one of the $W$ bosons decays leptonically. For the signal we use a $K$-factor of $2$ as at low values of $\tan \beta$ the $K$-factor for heavy Higgs production is similar to that of the SM Higgs \cite{Spira:1993bb,Muhlleitner:2006wx}.

Parton showering is modelled by \textsc{Pythia 8.2} \cite{Sjostrand:2014zea} and detector response by \textsc{Delphes3} \cite{deFavereau:2013fsa}. We have implemented the cuts to \textsc{MadAnalysis5 v1.7} \cite{Conte:2012fm,Conte:2014zja}. We also tested our signal benchmarks against the \textsc{MadAnalysis5} recast \cite{cmsrecast} of the CMS multilepton search \cite{Sirunyan:2017lae}, which is available at the \textsc{MadAnalysis} public analysis database \cite{Dumont:2014tja}. We were unable to exclude any of our BPs with the cuts of that analysis, we reached only $75\%$ CL exclusion at best.

We give a cutflow listing for signal and background events for the benchmarks discussed above in Tables \ref{tb:cutflow1} and \ref{tb:cutflow2} for SR1 ans SR2, respectively. For the background the dominant systematic uncertainty comes from the uncertainty of its shape as we have only rescaled the distribution from the LO one. In addition, the K-factors have uncertainties due to \textit{e.g.} scale variation.

In addition, the statistical uncertainty coming from the size of the MC sample is not negligible, this is especially the case for non-prompt leptons. As the cross section for $pp\rightarrow t\overline{t}$ is large, even a sizable sample will have event weights larger than one. The number of events passing our selection would be so small, that the error would be close to $30\%$, which would be too large compared to the expected level of systematics in this type of an analysis. We improved the statistics for non-prompt leptons by reverting the $b$ veto  and rescaling the cross section\footnote{At the \textsc{Delphes} level we used only a simple flat $85\%$ probability for $b$-tagging efficiency so this should not affect the kinematical cuts.}, which leads to an error below $15\%$, comparable to other missing factors and similar to the expected systematic errors\footnote{The errors are composed of uncertainties in trigger and tagging efficiencies, parton distribution functions and background estimates. The background errors arise either from shape errors and missing higher orders, if the estimate is based on simulation or from limited statistics and experimental uncertainties, if the background is estimated in a data-driven way. Typical error budgets for these kinds of final state topologies can be found in \cite{Khachatryan:2016kod,Sirunyan:2017uyt}.}. In an actual experiment the contribution from non-prompt leptons can be estimated in a data-driven way \cite{Chatrchyan:2011wba}.

\begin{table}
\begin{tabular}{l|c c c c|c c c c c c}
\hline
\hline
Cut & $WZ$ & $W^{\pm}W^{\pm}$ & non-prompt & Total bgnd & BP1 & BP2 & BP3 & BP4 & BP5 & BP6\\
\hline
Two SSSF leptons & $11625$ & $465$ & $10349$ & $22439$ & 305 & $42$ & $354$ & $479$ & $199$ & $361$\\
$p_{T}(\ell_{1})> 25$~GeV & $11507$ & $448$ & $9536$ & $21491$ & $294$ & $40$ & $343$ & $469$ & $158$ & $177$\\
$p_{T}(\ell_{2})> 12$~GeV & $11442$ & $431$ & $8530$ & $20403$ & $287$ & $40$ & $338$ & $461$ & $150$ & $166$\\
$Z$ veto & $4604$ & $431$ & $7630$ & $12665$ & $287$ & $39$ & $337$ & $459$ & $150$ & $166$\\
$p_{T}(\ell_{3})< 20$~GeV & $4395$ & $431$ & $6759$ & $11585$ & $286$ & $39$ & $337$ & $459$ & $150$ & $165$\\
\hline
$\slashed{E}_{T}> 50$~GeV & $1825$ & $295$ & $3854$ & $5974$ & $200$ & $28$ & $246$ & $322$ & $97$ & $112$\\
$M(\ell_{1}\ell_{2})\in[10,50]$~GeV & $321$ & $33$ & $992$ & $1346$ & $82$ & $8.7$ & $84$ & $108$ & $52$ & $87$\\
$b$-jet veto & $315$ & $28$ & $371$ & $714$ & $76$ & $7.7$ & $77$ & $98$ & $46$ & $80$\\
\hline
$M_{T}(\ell_{2})>100$~GeV & $7.7$ & $1.3$ & $13.3$ & $22.3$ & $20.0$ & $1.1$ & $23.8$ & $36.7$ & $4.6$ & $6.1$\\
\hline
\hline
\end{tabular}
\caption{We represent the cutflow for signal benchmarks and the background using the cuts of SR1 for an integrated luminosity of $137$~fb$^{-1}$ corresponding to LHC Run II.\label{tb:cutflow1}}
\end{table}

\begin{table}
\begin{tabular}{l|c c c c|c c c c c c}
\hline
\hline
Cut & $WZ$ & $W^{\pm}W^{\pm}$ & non-prompt & Total bgnd & BP1 & BP2 & BP3 & BP4 & BP5 & BP6\\
\hline
$\slashed{E}_{T}> 100$~GeV & $273$ & $131$ & $1035$ & $1439$ & $94$ & $14.5$ & $115$ & $160$ & $52$ & $70$\\
$M(\ell_{1}\ell_{2})\in[10,80]$~GeV & $74$ & $30$ & $600$ & $704$ & $82$ & $10.4$ & $85$ & $115$ & $49$ & $66$\\
$b$-jet veto & $71$ & $27$ & $186$ & $284$ & $72$ & $9.3$ & $76$ & $101$ & $43$ & $61$\\
\hline
$M_{T}(\ell_{2})>100$~GeV & $8.3$ & $3.3$ & $30.9$ & $42.5$ & $34.1$ & $2.9$ & $34.6$ & $50.4$ & $12.2$ & $10.1$\\
\hline
\hline
\end{tabular}
\caption{The cutflow for signal benchmarks and the background using the cuts of SR2. We only show the part of the cutflow that is different from Table \ref{tb:cutflow1}.\label{tb:cutflow2}}
\end{table}

In Table \ref{tb:siglist} we show the expected significances of the signal for the integrated luminosity of Run II with two assumptions on the total systematic uncertainties of $30\%$ and $20\%$. In general the signal regions show rather similar behaviour, SR1 being usually better when there is a compressed spectrum and SR2 better at the heavier end of the spectrum. In general there is a good chance of seeing a signal if there is reasonably much phase space for both decays while still having the heavy Higgs lighter than $500$~GeV. \pagebreak[4] As SR1 has the smaller SM background, it tends to perform better when there is a large number of overall signal events and hence it gives the larger expected significance also for BP4, where we might expect SR2 to perform better based on the spectrum. We see that for three of our signal benchmarks we would see a significant deviation from the SM background as long as the systematics are not too large.

\begin{table}
\begin{tabular}{c c c c c}
\hline
\hline
Benchmark & SR1, $30\%$ syst. & SR1, $20\%$ syst. & SR2, $30\%$ syst. & SR2, $20\%$ syst.\\
\hline
BP1 & $2.4\sigma$ & $3.1\sigma$ & $2.4\sigma$ & $3.2\sigma$\\
BP2 & $0.13\sigma$ & $0.17\sigma$ & $0.20\sigma$ & $0.27\sigma$\\
BP3 & $2.9\sigma$ & $3.7\sigma$ & $2.4\sigma$ & $3.2\sigma$\\
BP4 & $4.5\sigma$ & $5.6\sigma$ & $3.5\sigma$ & $4.7\sigma$\\
BP5 & $0.56\sigma$ & $0.71\sigma$ & $0.85\sigma$ & $1.1\sigma$\\
BP6 & $0.74\sigma$ & $0.94\sigma$ & $0.70\sigma$ & $0.94\sigma$\\
\hline
\hline
\end{tabular}
\caption{The expected statistical significances for the signal benchmarks for an integrated luminosity of $137$~fb$^{-1}$ assuming a total systematic uncertainty of $30\%$ or $20\%$. \label{tb:siglist}}
\end{table}

\begin{figure}
\includegraphics[width=0.46\textwidth]{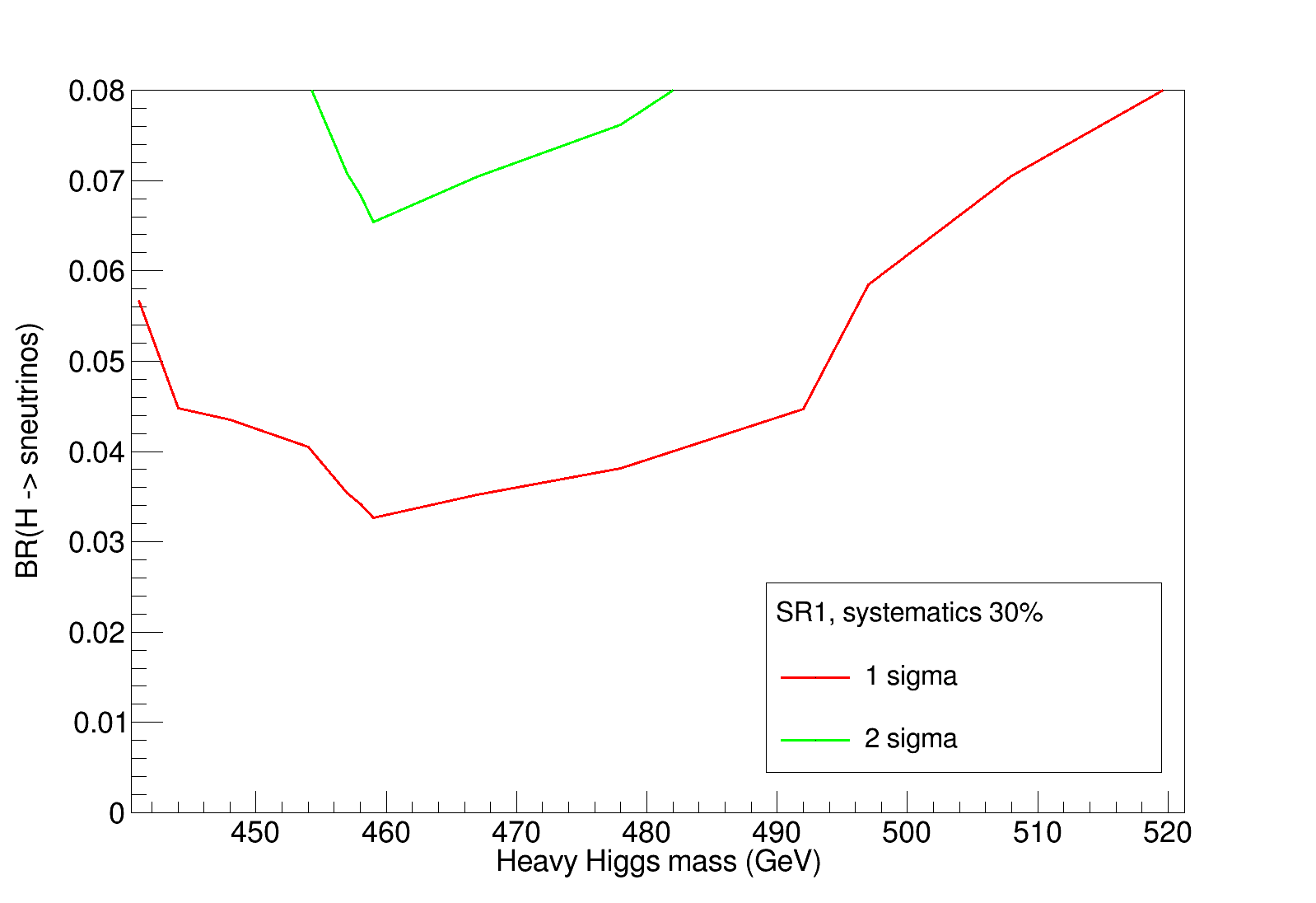}
\includegraphics[width=0.46\textwidth]{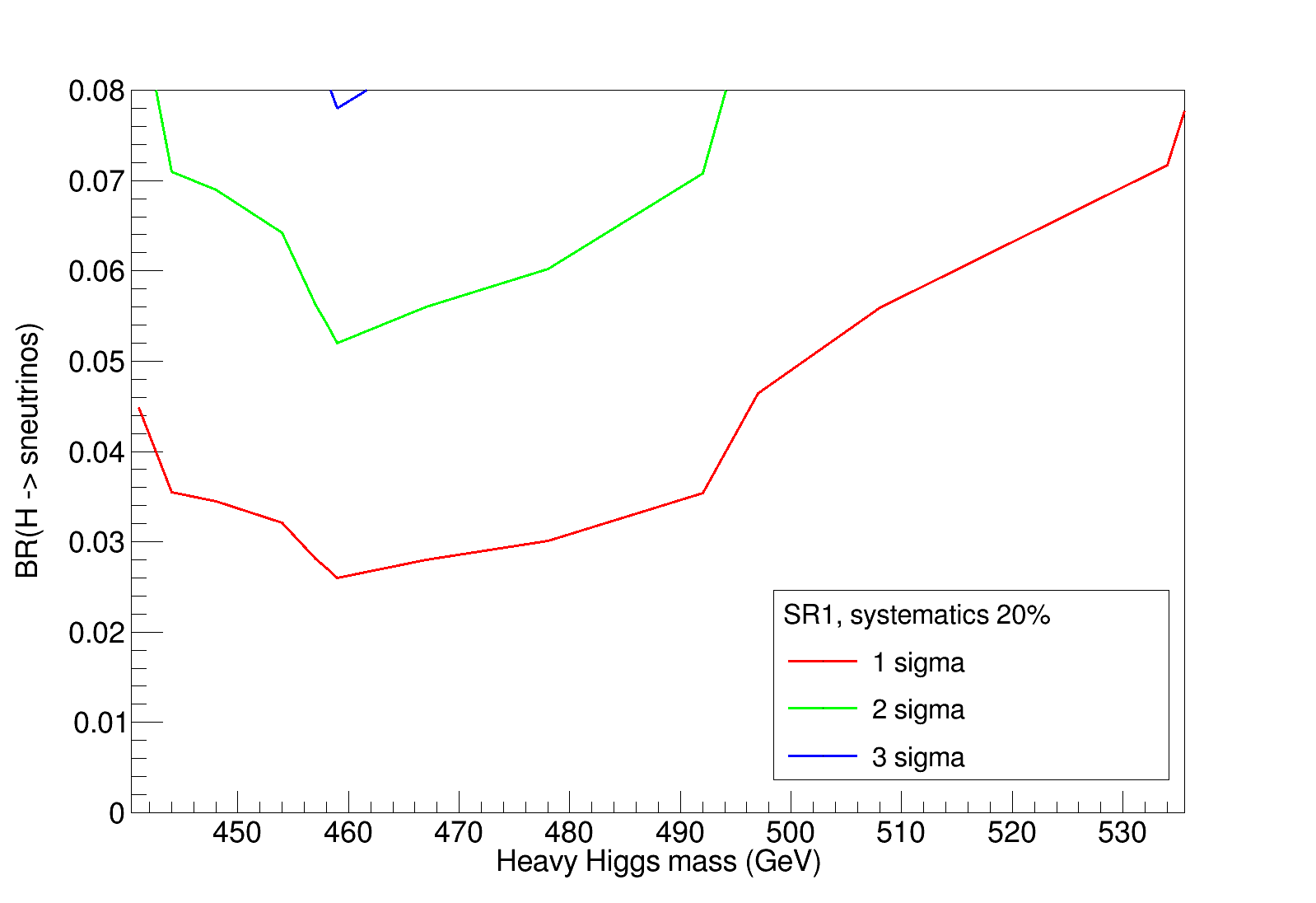}
\caption{The expected significances for SR1 assuming a total systematic error of $30\%$ or $20\%$. The parameters for this scan are given in Table \ref{tb:scanparams}. The kink around $455$~GeV is due to the CP-odd Higgs starting to contribute to the signal. \label{fig:SR1}}
\end{figure}

\begin{figure}
\includegraphics[width=0.46\textwidth]{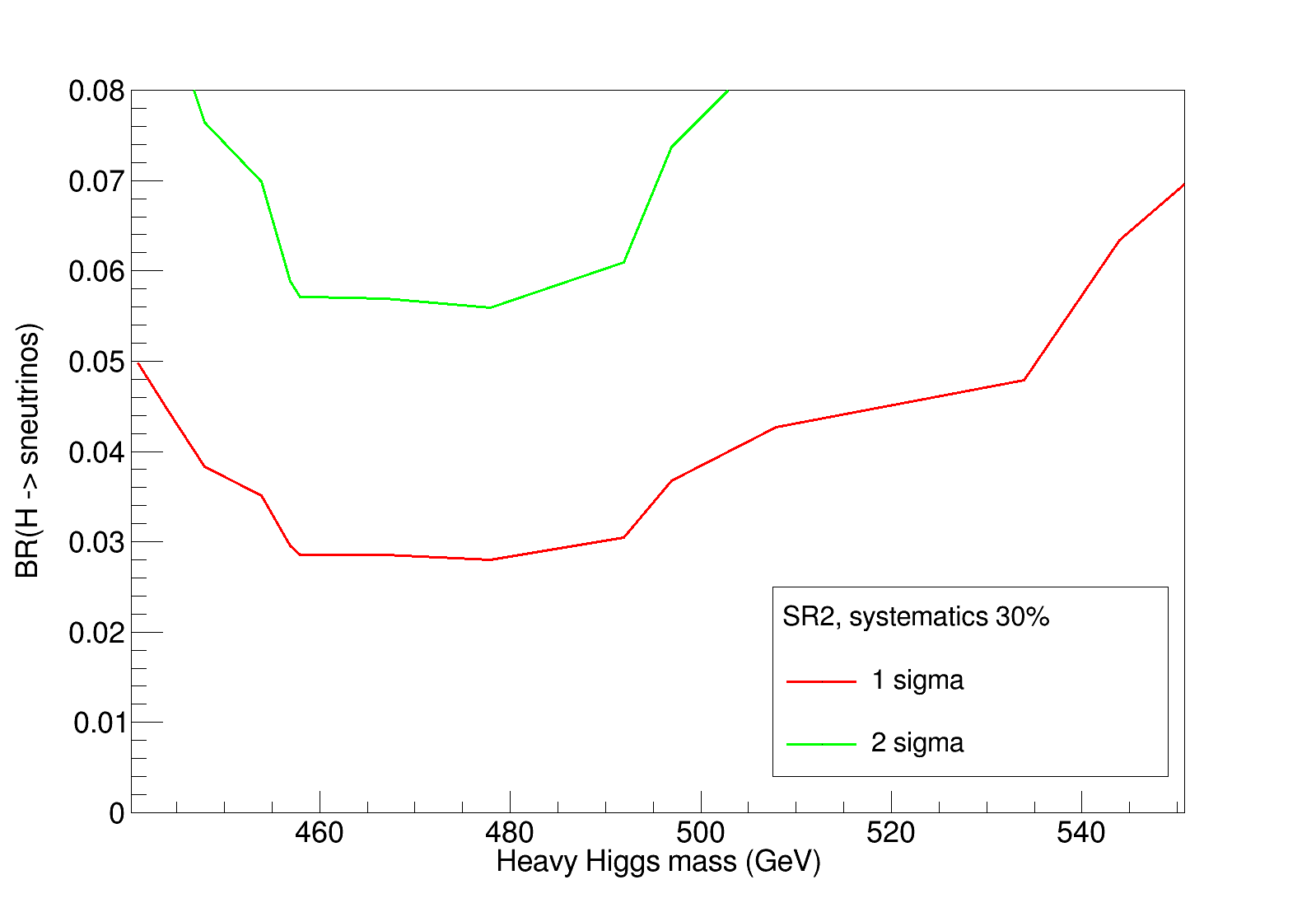}
\includegraphics[width=0.46\textwidth]{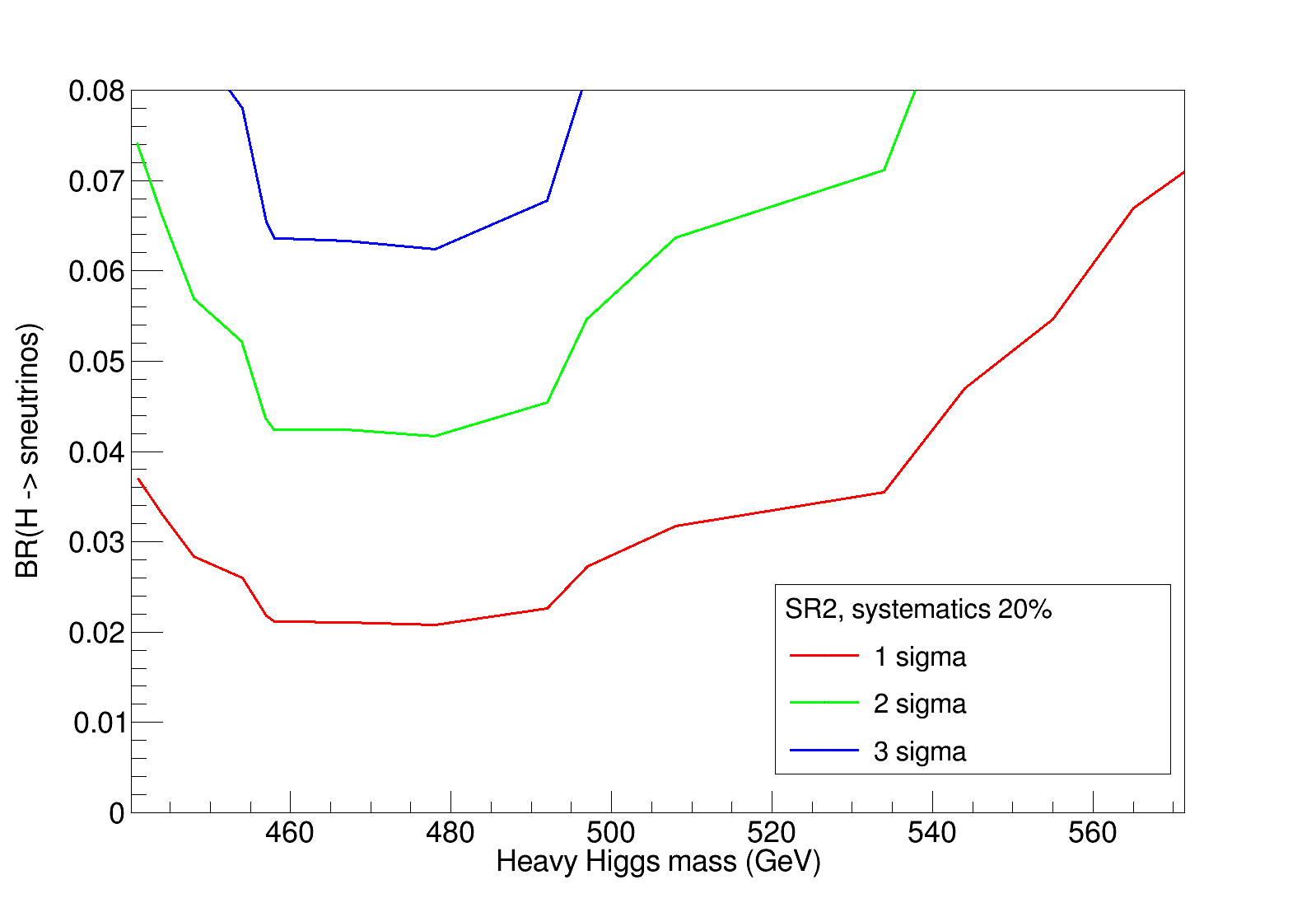}
\caption{The expected significances for SR2 assuming a total systematic error of $30\%$ or $20\%$. The parameters for this scan are given in Table \ref{tb:scanparams}. Notice that the range of masses is larger than in the plots of Figure \ref{fig:SR1}. \label{fig:SR2}}
\end{figure}

We also illustrate the dependence on the heavy Higgs mass holding the sneutrino and chargino masses fixed in Figures \ref{fig:SR1} and \ref{fig:SR2}. We give the most important parameters of this scan in Table \ref{tb:scanparams}. The signal region SR1 has its best sensitivity when $m_{A}$ is slightly above $2m_{\tilde{N}}$, while SR2 is sensitive to a larger range of masses. Above $m_{H}=500$~GeV the production cross section of the heavy Higgs starts to fall quite rapidly and only SR2 has limited sensitivity to the parameter space where the BR($H\rightarrow \tilde{N}\tilde{N}$) is large.

\begin{table}
\begin{tabular}{c c|c c}
\hline
\hline
Parameter & Value & Parameter & Value\\
\hline
$\tan \beta$ & $2.5$ & $\lambda$ & $0.52$\\
$\lambda_{N}$ & $0.55$ & $m(\tilde{\chi}^{\pm}_{1})$ & $186$~GeV\\
$m(\tilde{N}_{1})$ & $220$~GeV & BR($\tilde{N}\rightarrow \mu^{\pm}\tilde{\chi}^{\mp}$) & $37\%$\\
BR($\tilde{N}\rightarrow e^{\pm}\tilde{\chi}^{\mp}$) & $4\%$ &BR($\tilde{N}\rightarrow \tau^{\pm}\tilde{\chi}^{\mp}$) & $7\%$\\
\hline
\hline
\end{tabular}
\caption{The most important fixed parameters of the scan for plots of Figures \ref{fig:SR1} and \ref{fig:SR2}. We kept these parameters fixed and prepared a set of benchmarks with varying heavy Higgs masses and then simply rescaled the signal strengths according to the BRs.\label{tb:scanparams}}
\end{table}

The uncertainties are dominated by systematics if they are at the level of $30\%$, while if the systematics are at $20\%$ level the statistical and systematic errors are comparable in size and further data will improve the expected significances.

If the Center-of-Mass (CM) energy for Run III will be $\sqrt{s}=14$~TeV, the signal will be enhanced by roughly $20\%$, while the enhancement for the backgrounds is $10$--$15\%$ and hence the signal-to-background ratio will improve above that of Run II. The cut effiiencies will change slightly and it seems that those of SR2 are more robust against the increase of the CM energy while for SR1 background rejection is not as good as with $13$~TeV. Further improvements in $b$-tagging algorithms may also help to reject the background from $t\overline{t}$-originated non-prompt leptons. The question of how much the increase in statistics will help will depend on the final level of systematic errors. If they can be pushed below $20\%$, the increase in statistics will definitely help.

We must also note that the overall decay width depends on the absolute mass of neutrinos (for further discussion, see \cite{Arbelaez:2019cmj}). If the lightest neutrino is nearly massless, the Yukawa coupling is very small. Since the same coupling is responsible for the decay of the corresponding sneutrino to a lepton and a higgsino and, instead of a prompt decay, we may have displaced vertices with either a lepton and soft jets or two opposite-sign leptons plus missing transverse momentum in both cases. Typically one expects decay lengths of the order of a millimeter or so. Macroscopic decay lenghts would require one of the sneutrinos to be aligned in flavor space with the lightest neutrino so that the effective Yukawa coupling would be almost zero.

\section{Access to the neutrino mass generation mechanism}

Besides allowing a visible signal of lepton number violation, our model is interesting also in the sense that it allows us to probe the mechanism of neutrino mass generation. We shall now assume that we would at some point see a statistically significant excess of the type described and discuss how to use the information to constrain neutrino physics. A reasonably accurate measurement of the parameters will require the data from the HL-LHC stage.

\begin{figure}
\begin{center}
\includegraphics[width=0.45\textwidth]{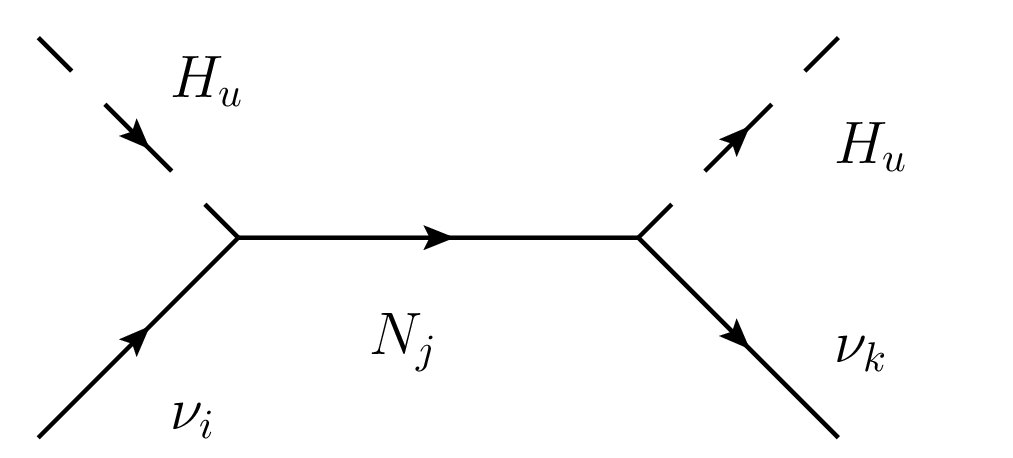}
\end{center}
\caption{The Feynman diagram responsible for the light neutrino masses in type-I seesaw.\label{fig:numass}}
\end{figure}

In the type-I seesaw mechanism the neutrino masses are based on the diagram in Figure \ref{fig:numass}. The light neutrino mass matrix elements will be
\begin{equation}\label{eq:numatrix}
m^{\nu}_{ik}=\sum_{j=1}^{3}\frac{y^{\nu}_{ij}y^{\nu\dagger}_{kj}v^{2}\sin^{2}\beta}{2m_{N_{j}}},
\end{equation}
where we have given the Yukawa couplings in a basis where the left-handed neutrinos are in the charged lepton flavor basis and the right-handed neutrinos in the basis of mass eigenstates. If either the right-handed neutrino masses are larger than the soft SUSY breaking masses for right-handed sneutrinos ($m_{N}^{2}\gg m_{\tilde{N}, \mathrm{soft}}^{2}$) or the soft SUSY breaking masses are aligned with the right-handed neutrino masses, these are the neutrino Yukawa couplings given in equation (\ref{eq:superpotential}). In the case where the heavy Higgs can decay to sneutrinos with a reasonably large rate, the right-handed neutrino masses are at least slightly larger than the soft SUSY breaking  sneutrino masses so we shall make this approximation in what follows.

Since the neutrino mass matrix is given in the flavor basis, we already know that this matrix can be diagonalized by the Pontecorvo--Maki--Nakagawa--Sakata (PMNS) matrix \cite{Maki:1962mu}, which mixing angles have been determined from neutrino oscillation experiments.

We shall first assume that only one sneutrino will be kinematically accessible, \textit{i.e.}, lighter than $m_{H}/2$. The sneutrino decays to same-sign dileptons give us a handle on neutrino Yukawa couplings. The BRs of $\tilde{N}_{j}\rightarrow \tilde{\chi}^{\pm}\ell_{i}^{\mp}$ are proportional to $|y^{\nu}_{ij}|^{2}$. These BRs are most reliably estimated from the dilepton final states, where there is no chance of misidentification in which leptons originate from the sneutrino decay. We must also be able to get an estimate on the signal strength to the final state with hadronic taus to estimate the contribution emerging from leptonic tau decays to the electron and muon final states. Once all three decay modes have been identified, the ratios $|y^{\nu}_{ik}/y^{\nu}_{jk}|^{2}$ may be determined from the event rates for each lepton flavor corrected with signal efficiencies.

Limits on the absolute scale of the Yukawa couplings can also be obtained. If the decays result in secondary vertices, the decay length depends on $|y^{\nu}_{ij}|^{2}$, the square of the higgsino component of the chargino and the available phase space. An upper limit for the phase space and thus a lower limit for the scale of Yukawa couplings can be determined if we are able to estimate the chargino and the sneutrino masses. We do expect to see an excess from the heavy Higgs in some other channel, the most likely being $H\rightarrow \tau^{+}\tau^{-}$. Although the $\tau^{+}\tau^{-}$ final state contains neutrinos, it will be possible to get an estimate of $m_{H}$, which results in an upper bound for the sneutrino mass of $m_{\tilde{N}}\leq m_{H}/2$. The chargino mass has already some lower limits from direct searches \cite{Aaboud:2017leg,Sirunyan:2018iwl} and, if the higgsinos are light, we would expect a signal to emerge from searches for SUSY with compressed spectra.

An upper limit for the Yukawa couplings can be obtained by assuming that the soft SUSY breaking masses $m_{\tilde{N}, \mathrm{soft}}^{2}$ are not negative in which case the right-handed neutrinos are lighter than right-handed sneutrinos, \textit{i.e.} also lighter than $m_{H}/2$. Then the upper limit for $y^{\nu}_{ij}$ comes from the upper limit for neutrino masses, \textit{i.e.}, the contribution from a single right-handed neutrino should not saturate the neutrino mass bounds from cosmology, $\sum m_{\nu}< 0.12$~eV \cite{Aghanim:2018eyx}, though the bound will require some assumption on the value of $\tan \beta$. We should note that this constraint can be computed without the knowledge of the signs of the Yukawa couplings as $\sum m_{\nu}=\mathrm{Tr}(m^{\nu})=\sum_{i,j}|y^{\nu}_{ij}|^{2}v^{2}\sin^{2}\beta/2m_{N_{j}}$.

To give an idea of the range of Yukawa coupling limits we would get we take the heavy Higgs to be $460$~GeV, the lightest chargino $190$~GeV, the lightest neutralino $185$~GeV, assume $\tan \beta > 1$ (\textit{i.e.}, $\sin^{2}\beta>1/2$) and the lightest chargino and neutralino to be pure higgsinos. If we assume a single dominant Yukawa coupling, the upper bound for $|y^{\nu}|$ would be $1.4\times 10^{-6}$. The lower bound assuming a lifetime $c\tau<1$~mm would be $2\times10^{-7}$. Hence, in such a case, the Yukawas could be determined within an order of magnitude or so even with rather crude estimates. Of course if the decays are not prompt, the lower bound will be lower but the decay length then allows us to estimate the size of the Yukawa coupling, not just give it a limit.

Furthermore, note that these bounds could be improved if we were able to estimate the sneutrino mass better than just knowing  $m_{\tilde{N}}<m_{H}/2$ and also all new information on the absolute mass scale of neutrinos will help.

This is about as much as we can deduce from a single sneutrino. If the heavy Higgs can decay to two sneutrino flavors, the first problem is to distinguish them. If one of the sneutrinos decays promptly and the other has displaced vertices, or the decay lengths of the sneutrinos differ, this would be the best chance of distinguishing the two sneutrinos. If both decay promptly, the kinematical distributions will be rather similar as both sneutrinos emerge from the decays of the heavy Higgs and result in identical final states with a higgsino, leptons and either neutrinos or jets.

If the two sneutrinos can be distinguished, a similar exercise can be done to deduce the values of $|y^{\nu}_{ij}|$, with better bounds as the contribution from two flavors have to be below the current neutrino mass bounds. If all three sneutrinos could be found, one could even try to determine the relative signs of the Yukawa couplings and then estimate the right-handed neutrino masses, as $U^{-1}m^{\nu}U$ should be diagonal, where $U$ is the PMNS matrix and $m^{\nu}$ is given by equation (\ref{eq:numatrix}). If there is no solution that gives us (within reasonable error bars) a diagonal form, then either the right-handed neutrino and sneutrino mass bases differ significantly or there are other contributions to neutrino masses (\textit{e.g.}, the one coming from $\Delta m_{\tilde{N}}$ through the three-loop diagrams).

\section{Conclusions}

While the need for new physics  is clearly established by several flaws of the SM, the choice of SUSY as BSM paradigm  is 
per se not sufficient to remedy all of the SM shortcomings, primarily because it does not make any predictions on the origin of neutrino masses, possibly the most compelling evidence that the SM needs to be surpassed by a new theoretical framework.  Therefore, it is mandatory to supplement model realisations of  SUSY with a mechanism with neutrino mass generation. From this perspective, it becomes then intriguing to assess the possibilities of accessing the rather elusive  dynamics  of  neutrinos (which is tested in non-collider experiments) through its SUSY mirror image, that is, the dynamics of sneutrinos, which can potentially be probed at the LHC. We have made here a first step in this direction by studying the conditions under which lepton number violation can occur in heavy Higgs decays to sneutrinos and what ensuing signals may be accessible at the CERN machine, including --- in the presence of the latter --- sketching a procedure to constrain the size of the intervening neutrino Yukawas. 

The NMSSM extended with a type-I seesaw incorporating right-handed neutrinos offers a possibility to observe lepton number violation in the sneutrino sector without introducing a too large loop contribution to neutrino masses. If the right-handed sneutrinos have a mass difference, it is communicated either through tiny Yukawa couplings or diagrams with at least three loops so the CP-even and CP-odd states can have a reasonable mass difference. Further,  the right-handed sneutrino can only decay through its Yukawa couplings --- provided that the singlino is heavier --- so its decay width is very small. Hence the parameter $x=\Delta m_{\tilde{\nu}}/\Gamma_{\tilde{\nu}}\gg 1$, which then leads to lepton number violating signatures.

The right-handed sneutrinos can be pair produced through the heavy Higgs portal provided by our BSM scenario, provided the heavy Higgs is within the reach of the LHC. If the couplings $\lambda$ and $\lambda_{N}$ are large, the heavier Higgses have a reasonably large BR to sneutrinos. The sneutrinos themselves decay only through tiny Yukawa couplings to higgsinos and leptons, provided the decay mode to a singlino and a right-handed neutrino is kinematically forbidden. In such a case lepton-number violating sneutrino decays can be observed. 

The decay $ \tilde{N}_{R}\rightarrow \ell^{\pm}\tilde{H}^{\mp}$ is especially interesting as it is visible. Since $\Delta m_{\tilde{N}}\gg \Gamma_{\tilde{N}}$, the sneutrino does not have a well defined lepton number so the decays may violate it. If both of the sneutrinos decay visibly, we may get a same-sign dilepton signature, for which the SM background is a lot smaller than for opposite-sign dileptons. By using the features of the signal (only soft objects beyond the two leptons, missing transverse momentum from neutralinos \textit{etc.}) we were able to build signal regions, which could allow us to see an excess compared to the SM background if the heavy Higgs and sneutrinos are within the reach of the LHC.

By adopting several BPs in our BSM theoretical scenario, each exemplifying a different SUSY spectrum of masses and couplings in the (s)neutrino sector yet all capturing heavy Higgs masses in the 450 to 550 GeV range, the only one accessible at the  LHC in our case, we have proven that already Run II presents a level of sensitivity to our signatures that in our opinion warrants further experimental investigation.

\section*{Acknowledgments}

SM is supported in part through the NExT Institute and the STFC consolidated Grant ST/L000296/1. SM and HW acknowledge the H2020-MSCA-RISE-2014 grant no. 645722 (NonMinimalHiggs). HW acknowledges the support from Magnus Ehrnrooth Foundation and STFC Rutherford International Fellowship (funded through MSCA-COFUND-FP, grant number 665593).

\end{document}